**Non-reactive sintering enhances density and ionic conductivity of NASICON solid electrolytes**

*Andrea Cornelio[1,2], Björn Mieller[3], Johannes Baller[2], Andrea Fantin[3], Janina Roik[4], Jessica Kindt[1], Jonas Krug von Nidda[1], Tim-Patrick Fellinger[1], Gustav Graeber[1,2,*]*

* gustav.graeber@hu-berlin.de

[1]Department 3: Containment Systems for Dangerous Goods; Energy Storage, Federal Institute for Materials Research and Testing (BAM), 12205 Berlin, Germany

[2]Department of Chemistry, Humboldt-Universität zu Berlin, 12489 Berlin, Germany

[3]Department 5: Materials Engineering, Federal Institute for Materials Research and Testing (BAM), 12205 Berlin, Germany

[4]Department 1: Analytical Chemistry; Reference Materials, Federal Institute for Materials Research and Testing (BAM), 12205 Berlin, Germany

Keywords: NASICON, sodium solid-state batteries, solid-state reaction, calcination, reactive sintering, ionic conductivity, cycling stability

## Abstract

NASICON materials are promising solid electrolytes for room-temperature sodium solid-state batteries and are typically synthesized via solid-state reaction. While sintering has been extensively studied, the effect of calcination on electrolyte properties remains poorly understood. In this work, the temperatures at which the NASICON phase forms in $Na_3Zr_2Si_2PO_{12}$ and $Na_{3.4}Zr_2Si_{2.4}P_{0.6}O_{12}$ are identified. Calcination temperature is then varied between 900 °C and 1200 °C to obtain powders with different degrees of reaction prior to sintering. Under identical sintering conditions, higher NASICON phase content in the calcined powder is shown to yield denser electrolytes. Non-reactive sintering also improves grain boundary conductivity, increasing it by 130% for $Na_{3.4}Zr_2Si_{2.4}P_{0.6}O_{12}$ and raising total conductivity from 1.60 to 2.95 mS·cm$^{-1}$ for powders calcined at 900 °C and 1200 °C, respectively. Finally, it is shown that phosphorus loss during processing compromises cycling stability against Na metal electrodes, and that adding off-stoichiometric phosphorus resolves this issue, while reaching a critical current density of 5.0 mA·cm$^{-2}$ and a room-temperature conductivity of 3.82 mS·cm$^{-1}$ with over 400 hours of stable cycling. Overall, these findings directly relate synthesis and processing conditions to the final material properties and electrochemical performance of NASICON solid electrolytes.

# 1. Introduction

Among new emerging battery technologies, sodium-based batteries have recently attracted broad research attention.[1] Sodium is abundant, cheap and widely distributed on Earth´s crust. It also has a very low reduction potential, -2.71 V *vs*. SHE, and a theoretical capacity of 1166 mAh·g$^{-1}$, properties that make this element promising for electrochemical applications.[2,3]

Solid-state sodium batteries can offer a safer and energy-dense alternative to state-of-the-art lithium-ion and sodium-ion batteries. The replacement of the flammable liquid electrolytes used in these technologies with inorganic solid electrolytes can in fact reduce safety hazards in electrical battery applications.[4] A further increase of both volumetric and gravimetric energy densities of the negative electrode can be achieved by directly employing metallic Na or energy-dense Na alloys as negative electrodes.[4–6]

For these reasons, research efforts towards the discovery and development of solid ionic conductors for room-temperature (RT) applications have sharply increased.[2,4,7] Among the various classes of sodium solid electrolytes (SEs), including sulfides, halides, hydrides, polymers, glasses and oxides, NASICON-type oxides have emerged as one of the most promising candidates due to their high ionic conductivity, almost-unitary transference numbers and excellent thermal and chemical stabilities.[2,4]

NASICON electrolytes for Na solid-state battery applications typically adopt the general composition $Na_{1+x}Zr_2(SiO_4)_x(PO_4)_{3-x}$ ($0 \leq x \leq 3$). The first and most extensively studied composition is $Na_3Zr_2Si_2PO_{12}$ (x = 2), since its introduction by Hong and Goodenough in 1976.[8-10] Recently, specialized studies have highlighted the great influence of the ratio between Si and P contents in the lattice on the ionic properties of these materials.[8,11] A Si/P ratio of 2.4/0.6 allows to greatly increase conductivity due to an optimal tradeoff between sodium vacancies and charge carrier concentration in the lattice.[8,11,12] The stoichiometry $Na_{3.4}Zr_2Si_{2.4}P_{0.6}O_{12}$ has therefore gained momentum as a new benchmark in recent years, reaching Na conductivities of up to 5 $mS \cdot cm^{-1}$ at RT.[11–13]

While optimization of the composition has substantially improved the transport properties of NASICON electrolytes, reproducibly achieving high performances in symmetric and full cells remains a significant challenge. Furthermore, nominally identical compositions frequently exhibit large variations in ionic conductivity, activation energy, critical current density, and cycling stability across the literature.[14,15] These changes originate from differences in ceramic processing, which governs phase composition, densification, grain growth, porosity, crystallinity, and grain boundary chemistry.[16–18] The resulting microstructure determines both the mechanical integrity and electrochemical performance of the final electrolyte.[19,20] Linking processing, structure and properties of the material is therefore essential for the rational design of high-performance NASICON SEs.

The solid-state reaction (SSR) route is one of the most widely employed methods for ceramic processing due to its simplicity, scalability, and compatibility with inexpensive, commercially available precursors. For NASICON solid electrolytes, SSR offers flexible control over composition and relies on cheap and readily accessible raw materials.[21,22]

The synthesis typically comprises two high-temperature steps: calcination and sintering. During calcination, volatile species are removed and solid-state reactions are initiated, leading to the formation of NASICON and secondary phases. Sintering subsequently promotes densification while phase evolution is completed. Depending on the choice of synthesis parameters, the NASICON phase may form during calcination, sintering, or both, with the reaction pathway and formation temperature strongly dependent on precursor chemistry and preparation. Although considerable effort has been devoted to optimizing sintering conditions to maximize densification and improve electrochemical properties of NASICON SEs,[13,23–28] the influence of the prior calcination step has received comparatively little attention. Peretti *et al.*[29] studied how an appropriate choice of calcination temperature can mitigate humidity effects on precursor materials and processing conditions. These results, obtained for the $Na_{3.4}Zr_2Si_{2.4}P_{0.6}O_{12}$ stoichiometry with SSR synthesis, also highlight how the NASICON phase is not fully formed until at least 1000 °C, while predominant NASICON phase appears in the samples calcined at 1230 °C. Xing *et al.*[30] compared SSR-produced $Na_3Zr_2Si_2PO_{12}$ materials with and without a calcination step at 900 °C before sintering, to explore the effect of a fully-reactive one-step sintering. NASICONs produced with fully-reactive sintering achieved good RT Na conductivity (1.48 mS·cm$^{-1}$) but a relatively low relative density (~90%). Wei *et al.*[31] synthesized $Na_3Zr_2Si_2PO_{12}$ and $Na_{3.2}Zr_2Si_{2.2}P_{0.8}O_{12}$ with the SSR route, exploring the relationship between calcined material composition in $Na_{3.2}Zr_2Si_{2.2}P_{0.8}O_{12}$ and the formation of a highly-conductive amorphous layer enhancing grain-boundary conductivity in the final SE. Finally, Li *et al.*[28] showed the effect of the choice of raw materials on NASICON phase formation at 1000 °C and on the final $ZrO_2$ impurity in the SE material. Due to the lack of systematic studies on the subject, no consensus exists regarding the appropriate calcination conditions, with reported temperatures ranging from roughly 900 to 1200 °C (**Figure 1**). Consequently, some studies perform sintering on fully reacted NASICON powders, while others rely on reactive sintering in which phase formation and densification occur simultaneously. This distinction is expected to be critical because the extent of phase evolution prior to sintering strongly influences densification kinetics as well as the microstructure and electrochemical performance of the electrolyte.[30,32]

This work systematically investigates the influence of calcination on the synthesis and performance of NASICON SEs. As introduced above, the two compositions $Na_3Zr_2Si_2PO_{12}$ and $Na_{3.4}Zr_2Si_{2.4}P_{0.6}O_{12}$ are arguably the most relevant for Na solid electrolyte applications and are therefore considered in this work. First, the thermal behavior and phase evolution of the two different stoichiometries are compared to study their distinct reaction pathways during solid-state synthesis. Then, the impact of incomplete NASICON phase formation on the densification behavior during sintering is examined. To isolate this effect, the calcination conditions are varied to produce powders with different extents of reaction prior to sintering, while the sintering schedule is kept constant and coherent with literature sources. This approach allows to directly assess whether reactive sintering promotes densification or instead compromises the final microstructure and properties of the electrolyte. Finally, the electrochemical performance of the resulting materials is evaluated by comparing not only their ionic conductivity but also their chemical stability in contact with Na electrodes.

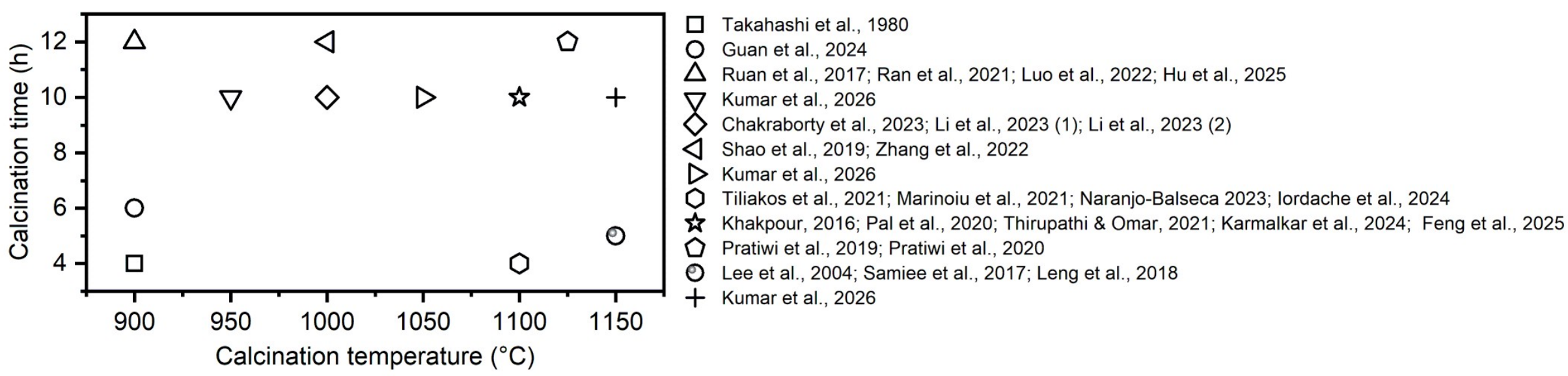

**Figure 1. Calcination temperatures and dwell times specified in literature for NASICON synthesis.** These calcination parameters all refer to $Na_3Zr_2Si_2PO_{12}$ NASICON synthesized by solid-state reaction (SSR) using $Na_2CO_3$, $ZrO_2$, $SiO_2$ and $NH_4H_2PO_4$ as precursor materials.[14,15,22,28,33-55]

# 2. Materials and methods

## 2.1 NASICON SSR synthesis

NASICON powders with compositions $Na_3Zr_2Si_2PO_{12}$ (referred to as $Na_3$) and $Na_{3.4}Zr_2Si_{2.4}P_{0.6}O_{12}$ (referred to as $Na_{3.4}$) are synthesized via a SSR route using $Na_2CO_3$ (Chemsolute, ≥ 99.8%), $SiO_2$ (abcr, 99.5%), $ZrO_2$ (Sigma-Aldrich, 99%), and $NH_4H_2PO_4$ (Chemsolute, ≥ 99%) as precursors. Additionally, $Na_{3.4}Zr_2Si_{2.4}P_{0.6}O_{12}$ with a 20% P excess ($Na_{3.4+0.2P}$) is also produced. This is done by increasing the molar amount of the P precursor by 20% in the mix of raw materials. Precursor particle sizes and quantities are reported in **Figure S1**. First, $NH_4H_2PO_4$ is pulverized in a vibration mill (Pulverisette 0, Fritsch), then all precursors are dried at 60 °C in a vacuum oven (≥ 24 h) and ball-milled in isopropanol (1:1 ml/$g_{powder}$) using a planetary ball mill (Pulverisette 6, Fritsch) with 50 Mg-stabilized $ZrO_2$ balls (10 mm, ball-to-powder ratio 2.5:1) for 60 min (3 × 10 min steps at 400 rpm followed by a 30 min step at 100 rpm, with 30 min pauses between each step). The pauses in-between steps allowed to avoid the overheating of the slurry and the milling jar, to add small and fixed amounts of isopropanol to keep the slurry´s texture constant, and to release ammonia vapors formed by partial $NH_4H_2PO_4$ decomposition. The final 100-rpm step proved to be effective in homogenizing the particle size of raw materials prior to calcination. After milling, the slurry is dried at 70 °C for ≥ 48 h, mixed in a Turbula mixer (20 min), and calcined in an open alumina cup crucible in air in a muffle furnace at 900, 1000, 1100, or 1200 °C for 12 h (heating rate 5 °C·min$^{-1}$).

The calcined agglomerates are crushed, pulverized, and milled again in isopropanol (1.5:1 ml/$g_{powder}$, ball-to-powder ratio 5:1) for 100 min at 300 rpm with 30 min pauses after every 30 min of milling (3 × 30 min steps and 1 × 10 min step at the end), followed by drying at 80 °C for ≥ 24 h. Pellets are uniaxially pressed at 150 MPa and sintered for 12 h at 1225 °C ($Na_3$) or 1300 °C ($Na_{3.4}$) with a heating rate of 5 °C·min$^{-1}$. The green bodies are sintered on a covered alumina substrate. To minimize Na and P volatilization during sintering and prevent reaction with the alumina support, the pellets are embedded in calcined mother powder. The sintered pellets are ground to ~1 mm thickness, polished with 4000-grit SiC paper, ultrasonically cleaned in ethanol, vacuum-dried, and annealed at 900 °C (heating rate 10 °C·min$^{-1}$). Annealing allows to reduce the negative effects of polishing and adsorbed $CO_2$ and $H_2O$ on the electrolyte surface prior to cell assembly.[56,57] The main steps of the synthesis procedure are schematized in **Figure 2**.

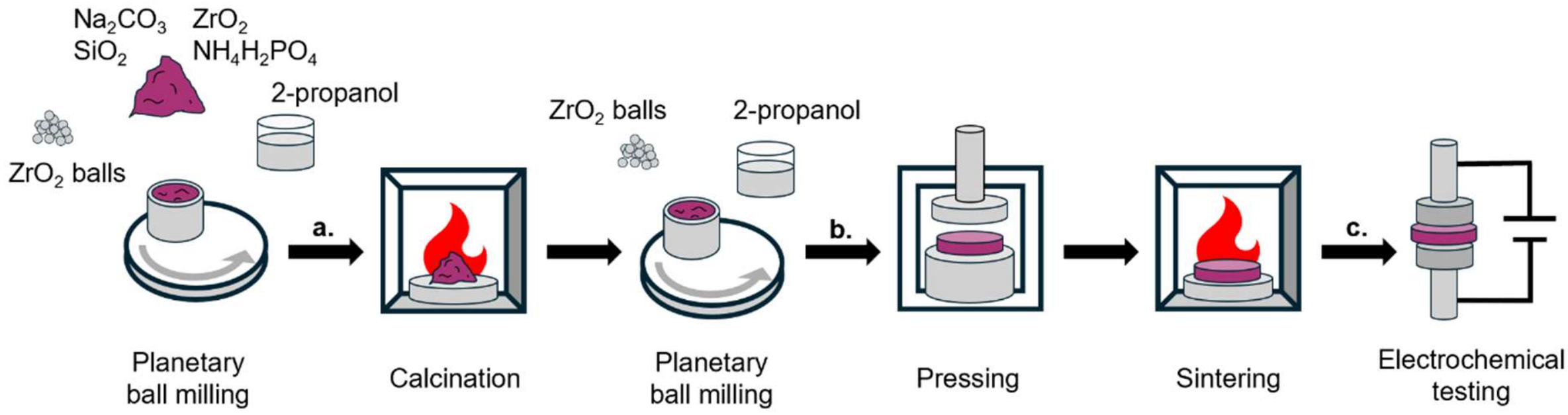

**Figure 2. Employed synthesis procedure and processing route.** The following characterization techniques are used at selected steps of the process: **a.** Particle size analysis (by laser diffraction), mass spectroscopy (MS), thermogravimetric analysis (TGA). **b.** Particle size analysis, thermal microscopy, X-ray diffraction (XRD). **c.** XRD, scanning electron microscopy (SEM), density measurements (Archimedes´ method in water), inductively coupled plasma optical emission spectrometry (ICP-OES).

## 2.2 Material characterization

Particle size distributions of raw materials, milled raw materials and milled calcined powders are analyzed with laser diffraction (Mastersizer 3000, Malvern Panalytical) in isopropanol.

Thermal decomposition of the precursor mixture is investigated by thermogravimetric analysis (TGA) (STA 449 F3 Jupiter, Netsch) in synthetic air coupled with mass spectroscopy (MS) (QMS 403 Quadro Aelolos, Netsch) up to 1450 °C with a 5 $°C·min^{-1}$ heating rate.

Calcined powders and sintered discs are analyzed with an X-ray diffractometer (D6 Phaser, Bruker) at room temperature with a Cu Kα X-ray source in reflection (Bragg-Brentano) configuration. A step of 0.005° between 4° and 80° ($2\theta$) is employed for calcined powers, and a step of 0.02° between 4° and 80° for sintered discs. Rietveld refinements are performed using TOPAS. The refinement models are initially developed in TOPAS-Academic V7 and the final refinements are carried out using TOPAS V6 (Bruker AXS, Karlsruhe, Germany).[58]

Thermal microscopy (EM301, Hesse Instruments) measurements are performed until 1450 °C (or melting of samples) with a 5 $°C·min^{-1}$ heating rate.

The relative density of the sintered pellets is measured with Archimedes´ method in deionized water.

Scanning electron microscopy (SEM) measurements on electrolytes are carried out with a FlexSEM II (Hitachi).

Inductively coupled plasma optical emission spectrometry (ICP-OES) is performed on dissolved sintered samples using an ARCOS ICP-OES analyzer (SPECTRO Analytical Instruments). Before measurement, 10-mg samples are weighed in quartz containers and dissolved using 20 ml of $H_2O$ and 5 ml of HCl on a hot plate at 150-180 °C for approximately 1 h. After dissolution and cooling, the samples are filtered into quartz flasks and topped up to a volume of 100 ml.

## 2.3 Electrochemical testing

Potentiostatic electrochemical impedance spectroscopy (PEIS) is performed from 5 MHz to 1 Hz using a BioLogic SP-200 potentiostat with an excitation voltage amplitude of ±15 mV. Conductivity measurements are performed on NASICON discs in blocking electrodes configuration in a piston cell (CompreCell 12, RHD Instruments) under an applied pressure of 25 MPa (CompreFrame, RHD Instruments). For this purpose, a 200-nm thick layer of gold (Polymet, min. 99.99% Au) is deposited on both sides of the discs using a thermal evaporator (CREAMET 300 Multi3, Creavac). Fitting of EIS spectra is performed using the open-source Python package DearEIS.[59]

Distribution of relaxation times (DRT) analysis on PEIS data is performed over the frequency range 5 MHz to 50 Hz using the TR-RBF method with the DearEIS software. Generalized cross-validation is used for the regularization parameter (λ). Inductive effects at high frequencies are not included in the analysis.

For critical current density and cycling tests, symmetrical Na | NASICON | Na cells are built in an Ar glovebox and tested in custom-made inset cells on which pressure is applied using an axial screw (a schematic of the custom-made cell is reported in **Figure S11**). A torque of 1 Nm is applied on the screw, resulting in ~10 MPa applied pressure. The diameter of the NASICON discs in contact with the Na electrode is 10 mm for all Na | NASICON | Na cells. Electrochemical measurements are performed using either a BioLogic SP-200 or a Gamry Reference 3000 potentiostat. Sodium electrodes are prepared in the glovebox by cutting, rolling and punching sodium discs from sodium cubes (Sigma-Aldrich, 99.9%). The electrodes´ surfaces are cleaned from oxidation products with a stainless steel spatula and a plastic brush right before assembly. Each electrode has a mass of approximately 20 mg.

All electrochemical measurements are performed at room temperature (20-25 °C) inside an Ar glovebox. No active temperature control is employed, but temperatures during each measurement are monitored using a thermocouple (type K) connected to a data logger.

# 3. Results and discussion

### 3.1 Synthesis and calcination of NASICONs

To investigate the influence of the initial phase composition on subsequent sintering behavior, powders exhibiting different degrees of reaction are produced by calcining the two NASICON compositions, $Na_3$ and $Na_{3.4}$, at different temperatures. To ensure that any differences between the samples are exclusively due to the calcination step, large batches are first prepared for each stoichiometry, which are then divided into identical sub-batches to limit the risk of having different particle size distributions between batches.

Thermal analysis is employed to identify the reaction sequence during heating and to determine the temperatures associated with NASICON formation (**Figures 3a-b**). Three distinct stages are defined during heating: (i) the decomposition of $NH_4H_2PO_4$ and $Na_2CO_3$, and the release of volatile species below 800 °C; (ii) the reactions leading to the formation of NASICON and its precursors without volatiles release; and (iii) the rapid crystallization of NASICON from its precursors between 1000 °C and 1200 °C, depending on the stoichiometry. These three stages are highlighted with colored boxes in **Figures 3a-b** for the two stoichiometries.

During stage (i), $NH_4H_2PO_4$ decomposes at around 200 °C, releasing $NH_3$ and $H_2O$ vapors (**Figures 3c-d**). The reaction with $Na_2CO_3$, during which $CO_2$ is released, begins at approximately 400 °C and continues until below 800 °C. The quantitative conversion of the carbonate is apparent by the absence of emitted $CO_2$ above 800 °C, corresponding to the thermal decomposition temperature of $Na_2CO_3$. Above 800 °C, no further mass loss is observed, indicating that volatile-producing reactions are complete. In stage (ii), the formation of NASICON and its precursors (*e.g.* $Na_2ZrSi_2O_7$, parakeldyshite) progresses without further mass loss. While NASICON phase formation can already be observed at these temperatures, it is not yet the prevalent phase in the mixture. Finally, in stage (iii), the endothermic event immediately followed by an exothermic peak suggests that NASICON crystallizes from a transient liquid phase (see red arrows in **Figures 3a-b**). This occurs between 1000 °C and 1100 °C for $Na_3$ and between 1100 °C and 1200 °C for $Na_{3.4}$. This distinct thermal signal coincides with a significant increase in NASICON content for both stoichiometries (**Figures 3e-h**). Melting of the NASICON phase occurs at approximately 1300 °C for $Na_3$ and 1400 °C for $Na_{3.4}$. The corresponding endothermic event is followed by an exothermic peak associated with NASICON decomposition into its constituent oxides. The final mass loss at temperatures above the melting point can be related to the evaporation of phosphorus, as detected by the MS measurements (**Figures 3c-d**). Furthermore, the total mass loss for both stoichiometries is lower than the expected theoretical mass loss (for $Na_3$, the expected total mass loss is ~17.2% of the raw powder's weight, while for $Na_{3.4}$ it is ~15.8%). The mass losses observed with TGA for the two stoichiometries are ~6.7% and ~10.5% for respectively $Na_3$ and $Na_{3.4}$, signifying that part of the volatile species are already released prior to the high-temperature reaction step. The ammonia vapors released during milling are in fact a signal of partial decomposition of reactants already before calcination.

$Na_3$ crystallizes at lower temperatures, while $Na_{3.4}$ NASICON forms at higher temperatures. This is consistent with the observations of Loutati *et al.*.[60] Overall, thermal analysis results demonstrate that the two stoichiometries follow similar reaction pathways but exhibit significantly different crystallization temperatures, providing the basis for selecting distinct calcination parameters.

Based on these results, the selected calcination temperatures (highlighted as vertical dashed lines in **Figures 3a-d**) are: 900 °C, 1000 °C, 1100 °C and 1200 °C (the latter only for $Na_{3.4}$). In all cases, the calcination temperature is held for 12 hours. To verify the expected evolution of phases, the calcined powders are subsequently milled and analyzed by XRD (**Figures 3e-f**). At lower temperatures, (almost) no NASICON phase forms. At intermediate temperatures, the secondary phases change and a limited amount of the NASICON phase is produced. At high temperatures (1100 °C for $Na_3$ and 1200 °C for $Na_{3.4}$), the crystalline portion of the calcined powders is predominantly NASICON. After calcination, the batches are named as follows: $Na_3^{900\,°C}$, $Na_3^{1000\,°C}$, $Na_3^{1100\,°C}$, $Na_{3.4}^{900\,°C}$, $Na_{3.4}^{1000\,°C}$, $Na_{3.4}^{1100\,°C}$, and $Na_{3.4}^{1200\,°C}$, where the subscript refers to the composition and the superscript to the calcination temperature.

Quantitative Rietveld refinements are performed for the expected crystalline phases. These results show the evolution of the crystalline phases with calcination temperature (**Figures 3g-h**). At low and intermediate temperatures, the powders consist mainly of $Na_2ZrSi_2O_7$, together with residual monoclinic $ZrO_2$ and minor amounts of other precursors. The NASICON phase, although starting to form at lower temperatures, becomes predominant for both stoichiometries at the highest calcination temperature, in line with the TGA results. For the $Na_3$ composition, the NASICON phase is 90.2 wt% after calcination at 1100 °C. The fractions of $Na_2ZrSi_2O_7$ and $ZrO_2$ are 5.2 and 4.6 wt%, respectively. The $Na_{3.4}$ reaction is delayed and takes place at higher temperatures. In this case, $Na_2ZrSi_2O_7$ remains the major phase up to 1100 °C. NASICON content highly increases after calcination at 1200 °C, with both monoclinic and rhombohedral polymorphs of NASICON appearing. The total NASICON content reaches 84.2 wt% at 1200 °C, with $Na_2ZrSi_2O_7$ (8.4 wt%), $ZrO_2$ (5.7 wt%), and $Na_3PO_4$ (1.7 wt%) remaining as secondary phases. The complete results of the refinements are reported in **Table S1**.

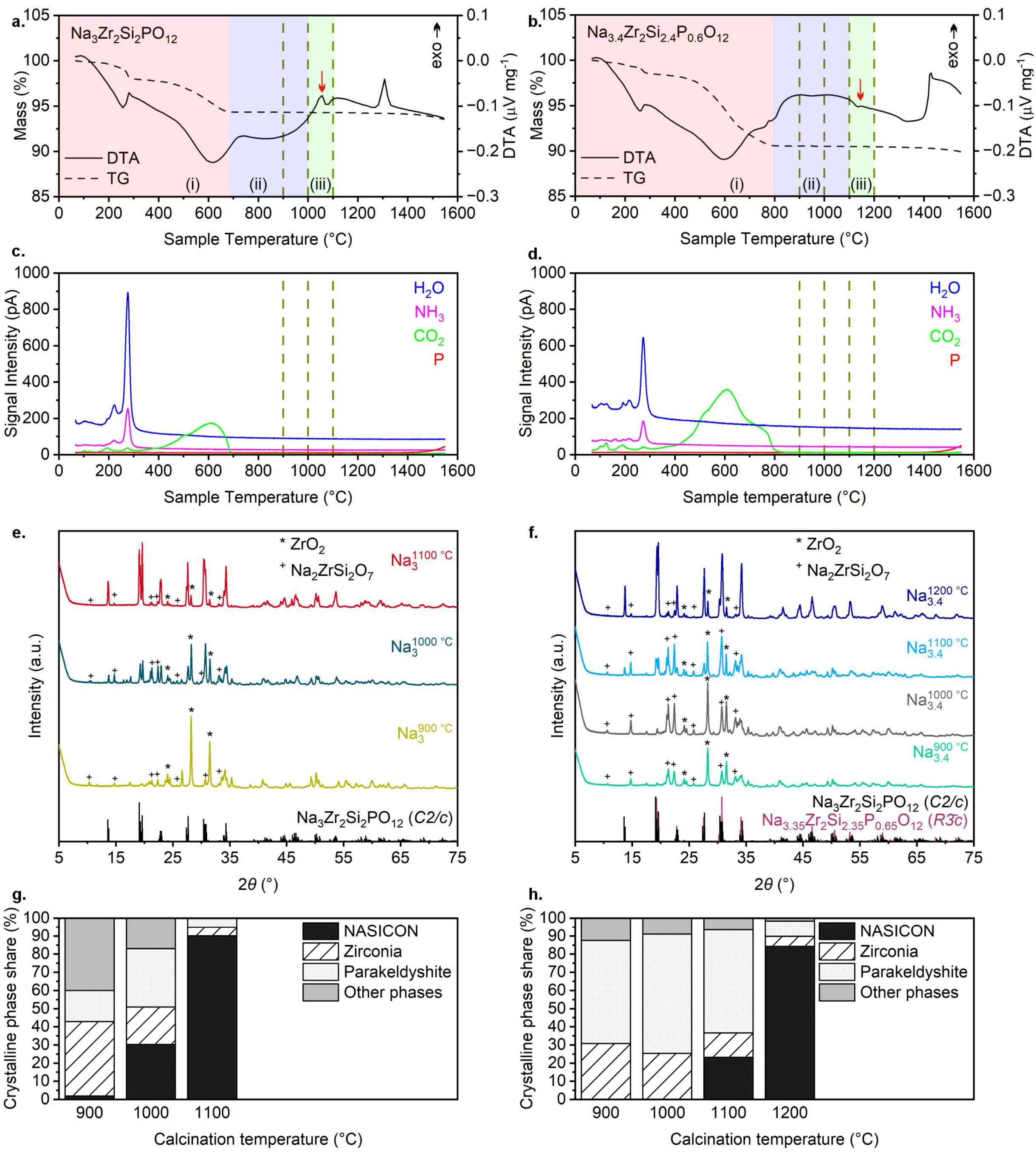

**Figure 3. Phase evolution of NASICON powders as a function of calcination temperature. a-b.** Thermogravimetric analysis (TGA) of milled precursor powders with nominal compositions $Na_3$ (**a.**) and $Na_{3.4}$ (**b.**), measured up to 1550 °C. Vertical dashed lines highlight the selected calcination temperatures for each composition. The three reaction stages during which high-temperature decomposition reactions start and volatiles are released (i), precursor and NASICON phases form without the release of volatiles (ii) and NASICON phases crystallize (iii) are highlighted with different colors. The red arrows in reaction stage (iii) point at the thermal signals

associated with NASICON phase formation. **c-d.** Mass spectroscopy (MS) results for $Na_3$ (**c.**) and $Na_{3.4}$ (**d.**), showing the temperature ranges during which volatiles are released due to raw material decomposition and subsequent chemical reactions. **e-f.** X-ray diffraction (XRD) of powders calcined at different temperatures. Measurements are performed after the second planetary milling step (refer to **Figure 2**). Peaks for $Na_3Zr_2Si_2PO_{12}$ (monoclinic, space group *C2/c*) and $Na_{3.35}Zr_2Si_{2.35}P_{0.65}O_{12}$ (rhombohedral, space group $R\overline{3}c$) refer to, respectively, entries #8096 and #62386 in the ICSD database, which are then used as a basis for refinement. The most representative peaks of monoclinic $ZrO_2$ and triclinic $Na_2ZrSi_2O_7$ phases are labelled. The sharp $Na_2ZrSi_2O_7$ peak at ~31° is not reported in the uppermost curves (highest calcination temperatures) due to its strong overlap with NASICON peaks. **g-h.** Crystalline phase fractions obtained with Rietveld refinement for all $Na_3$ (**g.**) and $Na_{3.4}$ (**h.**) batches calcined at different temperatures. Quantification refers exclusively to crystalline phases, as amorphous phases are not included in the refinement. Refinement curves are reported in **Figures S4-S5**, while the complete results are shown in **Table S1**.

### 3.2 Sintering and densification behavior

The influence of the initial phase composition on densification is first evaluated by heating microscopy. The shrinkage of the discs is computed from the frontal area detected by the camera and is reported *vs.* the sample temperature in **Figures 4a-b**. Although heating conditions differ from the actual sintering process because of the absence of mother powder, the different sample geometry and the lack of dwell time, heating microscopy provides a reliable comparison of the onset and progression of densification between the different powders. Both materials, $Na_3$ and $Na_{3.4}$, begin to shrink at approximately 1100 °C. However, $Na_{3.4}$ exhibits a lower densification rate and requires higher temperatures to achieve comparable shrinkage. As reported in previous literature,[31,60] densification takes place at higher temperatures for lower P content. This has been linked to the formation of P-containing glasses for higher P amounts. These glassy phases have a low melting point and favor reactions at lower temperatures.[60] Therefore, the $Na_{3.4}$ materials generally need to be sintered at higher temperatures. This shrinkage behavior mirrors the delayed crystallization of the $Na_{3.4}$ composition observed during thermal analysis, indicating that the differences in the initial phase formation persist during sintering and influence densification kinetics.

Based on the *in-situ* heating microscopy results, supported by previous studies and preliminary sintering trials, sintering temperatures of 1225 °C and 1300 °C are selected for $Na_3$ and $Na_{3.4}$, respectively.[48,50,56,61] These temperatures are selected to achieve comparable final densities while remaining below the onset of melting, thereby minimizing the risk of decomposition and Na and/or P volatilization during the sintering step. All samples are sintered for 12 h with the same protocol and in identical conditions. This is done to ensure that all observed differences in the resulting electrolytes can be related to their different calcination history. This way, the effect of calcination is isolated.

After sintering, Archimedes density measurements reveal a progressive increase in density with increasing calcination temperature, consistent with SEM observations showing a corresponding reduction in residual porosity (**Figure 4c**). $Na_3^{1100\,°C}$, $Na_{3.4}^{1100\,°C}$ and $Na_{3.4}^{1200\,°C}$ achieve the highest densities and the lowest residual porosities. These results indicate that the amount of NASICON phase formed prior to sintering strongly influences the subsequent densification behavior of NASICON electrolytes.

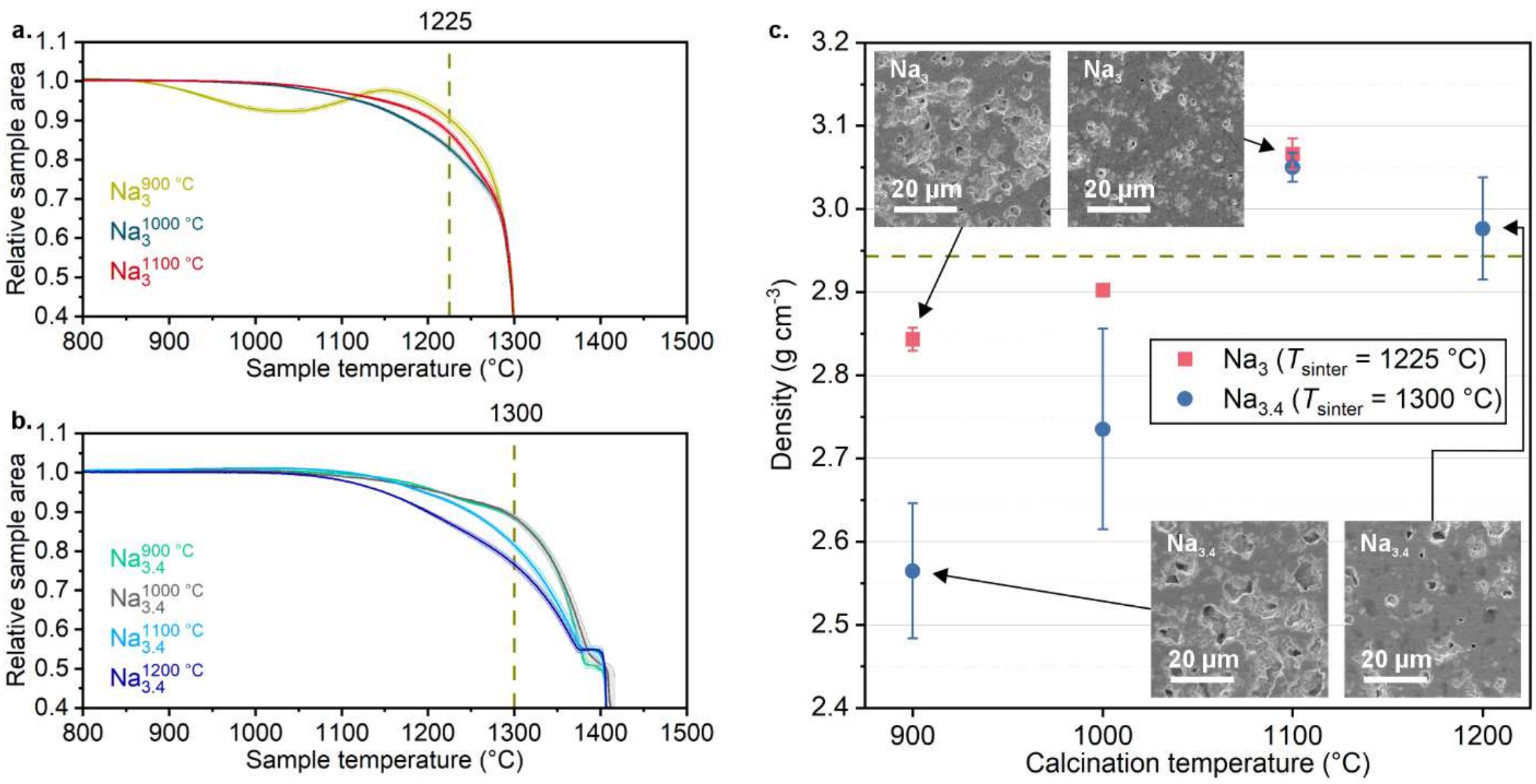

**Figure 4. Densification behavior of NASICON electrolytes during sintering depending on calcination temperature. a-b.** Normalized frontal area measured by thermal microscopy as a function of sample temperature up to the melting point. Each experiment is performed twice. Solid lines represent the average of the two measurements, while the shaded regions indicate the deviation between them. Vertical dashed lines indicate the selected sintering temperatures: 1225 °C for $Na_3$ samples and 1300 °C for $Na_{3.4}$ samples. Results are normalized to the sample frontal area measured at 600 °C to minimize artifacts occurring at lower temperatures, where no densification is expected. **c.** Raw density of sintered samples measured with the Archimedes´ method in water. 3 discs are measured for each stoichiometry. Results include error bars that show

the standard deviation between measurements. The horizontal dashed line corresponds to 90% of the theoretical density for the monoclinic $Na_3$ phase (3.27 $g{\cdot}cm^{-3}$).[19] Scanning electron microscopy (SEM) images of representative electrolytes (after polishing with 4000-grit SiC sandpaper) are included as insets. SEM images of all electrolytes are shown in **Figure S9**.

### 3.3 Phase and compositional evolution during processing

As-sintered electrolytes are analyzed by XRD to determine the final phase purity after sintering. All samples exhibited NASICON phase as the main crystalline phase, with monoclinic $ZrO_2$ as the main secondary phase (**Figures 5a-b**). Monoclinic $ZrO_2$ in sintered NASICONs produced with SSR is a common and well-known impurity phase.[28] The presence of the same phases across all calcination conditions indicates that the different initial phase compositions are converted into the same equilibrium phase composition during sintering. All electrolytes reach full phase conversion after the sintering step, regardless of the calcination history. To evaluate whether the different calcination histories result in variations in elemental composition after sintering, selected samples are analyzed by ICP-OES. This is done to verify whether differences in the extent of NASICON formation before sintering can influence volatile element losses during the subsequent thermal treatment.

ICP-OES results (**Figures 5c-d**; full data reported in **Table S2**) reveal similar elemental compositions among the different batches, indicating that the calcination history does not significantly affect the retention of Na, Zr, Si, or P during processing. However, all samples exhibit a systematic phosphorus depletion of approximately 20%. This loss is attributed to the volatilization of phosphorus-containing species (likely, phosphorous oxides) during calcination, which occurs before complete incorporation of phosphorus into the NASICON structure.

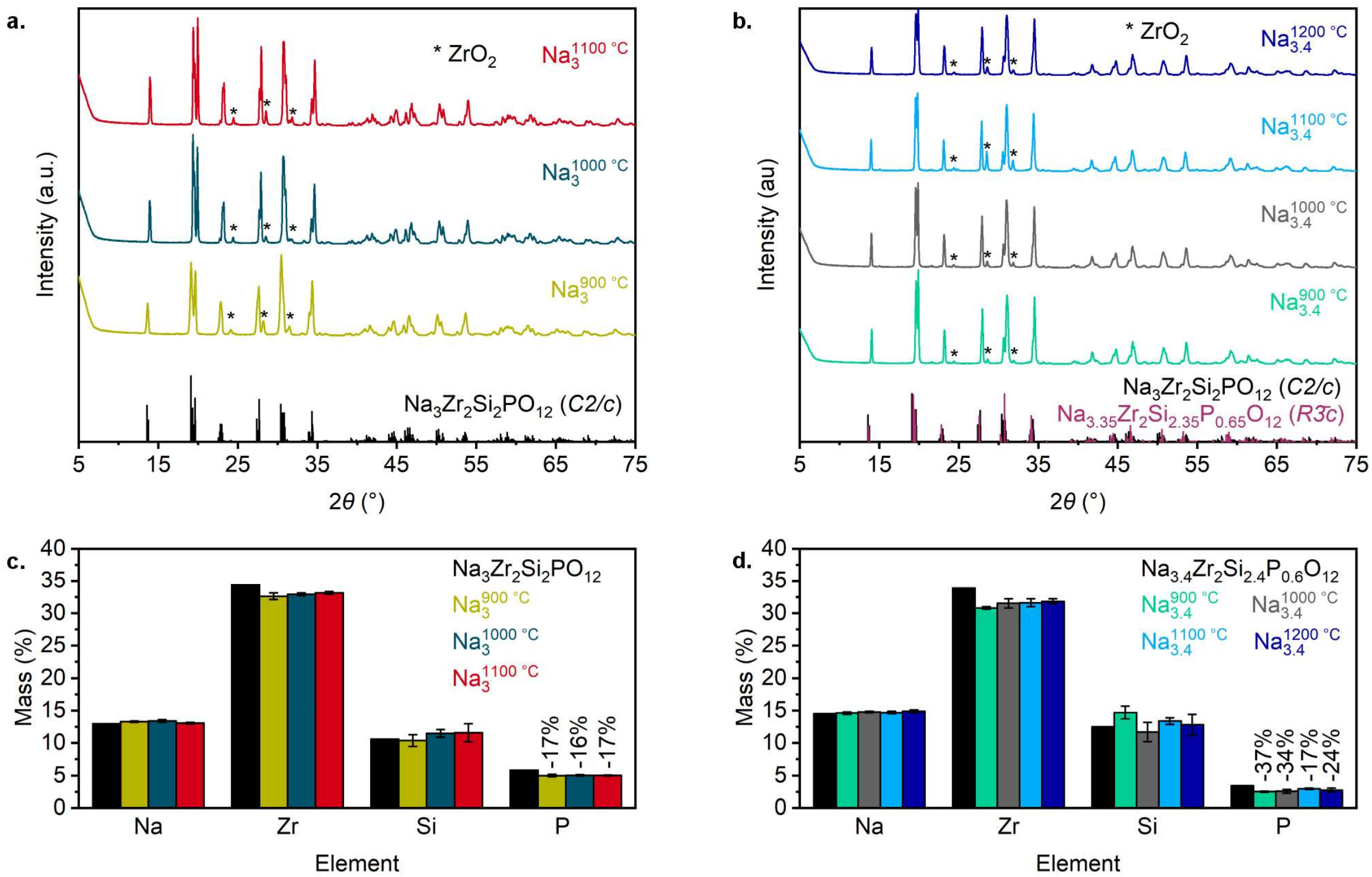

**Figure 5. Structural and compositional characterization of sintered electrolytes. a-b.** XRD results for the as-sintered discs for the investigated compositions. Peaks for $Na_3Zr_2Si_2PO_{12}$ (monoclinic, space group *C2/c*) and $Na_{3.35}Zr_2Si_{2.35}P_{0.65}O_{12}$ (rhombohedral, space group $R\overline{3}c$) refer to, respectively, entries #8096 and #62386 in the ICSD database. Characteristic peaks of the monoclinic $ZrO_2$ impurity phase are highlighted. **c-d.** Inductively coupled plasma optical emission spectroscopy (ICP-OES) analysis of the corresponding sintered electrolyte materials. For each composition, the nominal elemental concentrations calculated from the target stoichiometry are reported in the left-hand columns (in black). The error bars for the measured values are the standard deviation computed from three measurements. The percentage error for phosphorous content is also reported numerically, as phosphorous lack is observed in all samples and is thought to influence the final properties of the electrolyte material. The oscillations in Si content in the samples are most likely related to the higher deviation of these measurements. A systematic lack of Zr is observed in all samples, but its relative value is lower compared to the detected P loss (refer to **Table S2** for complete ICP-OES results).

### 3.4 Electrochemical properties of NASICON electrolytes

The influence of calcination history on ionic transport is evaluated by measuring the impedance response of all sintered electrolytes using blocking electrodes (**Figure 6**). Since calcination primarily alters the extent of phase formation prior to sintering, impedance spectroscopy can directly reveal whether the resulting differences in densification and microstructure translate into changes in ionic transport. Equivalent circuit fitting (**Figures 6a, 6c**) enables separation of bulk and grain-boundary contributions, allowing the effect of calcination on each transport process to be evaluated independently (**Figures 6b, 6d**). Bulk conductivity ($\sigma_{bulk}$) reflects the intrinsic transport properties of the crystal structure and is therefore mainly dictated by composition.[11,12] Grain-boundary conductivity ($\sigma_{gb}$), in contrast, is highly sensitive to microstructure, grain contacts, secondary phases and residual porosity.[12,19]

Bulk conductivity mainly changes between stoichiometries, with $Na_{3.4}$ electrolytes showing considerably higher bulk conductivity compared to the $Na_3$ series due to their improved ionic structure.[8] For both stoichiometries, $Na_3$ and $Na_{3.4}$, bulk conductivity monotonically increases with calcination temperature (refer to **Figure S12** for complete data on conductivity contributions). Although the intrinsic bulk transport is primarily determined by composition, this trend may reflect the reduced fraction of residual secondary phases or improved structural homogeneity and crystallinity achieved after calcination at higher temperatures. The effect of calcination on bulk ionic transport is smaller than on grain-boundary conductivity. In the case of $Na_3$ electrolytes, $\sigma_{bulk}$ increases 11% from the $Na_3^{900\ °C}$ sample to the $Na_3^{1100\ °C}$ sample, while $\sigma_{gb}$ increases by 78%. $\sigma_{bulk}$ and $\sigma_{gb}$ increase of 49% and 130% respectively between $Na_{3.4}^{900\ °C}$ and $Na_{3.4}^{1200\ °C}$.

The much stronger dependence of grain-boundary conductivity on calcination history indicates that it is microstructural evolution, rather than intrinsic crystal chemistry, that dominates the observed improvements. Higher calcination temperatures lead to more complete phase formation prior to sintering, resulting in higher density, improved grain contacts and fewer resistive grain-boundary regions. These effects substantially increase grain-boundary conductivity. Since the grain-boundary contribution dominates the total impedance of polycrystalline ceramics, improvements in grain-boundary transport directly translate into higher total conductivity.[12] Impedance spectra reveal a systematic increase in total conductivity with increasing calcination temperature, consistent with the increased density observed after sintering. In the $Na_3$ series, $Na_3^{1100\ °C}$ reaches a grain boundary conductivity of 2.95 $mS \cdot cm^{-1}$, while the counterparts calcined at lower temperatures, $Na_3^{900\ °C}$ and $Na_3^{1000\ °C}$, reach values of 1.66 and 1.77 $mS \cdot cm^{-1}$ respectively. This improvement largely accounts for the higher total conductivity of $Na_3^{1100\ °C}$. The same trend remains for $Na_{3.4}$ electrolytes, with $Na_{3.4}^{1200\ °C}$ reaching a grain boundary conductivity of 6.67 $mS \cdot cm^{-1}$, clearly surpassing the other electrolytes.

The electrolytes calcined at higher temperatures – respectively $Na_3^{1100\,°C}$ and $Na_{3.4}^{1200\,°C}$ for the two compositional groups – exhibit the highest bulk and grain-boundary conductivities. Their total conductivities are, respectively, 0.98 and 2.95 mS·cm$^{-1}$. All electrolytes exhibited $Na^+$ transference numbers exceeding 0.99, confirming essentially pure ionic conduction (DC polarization results for all electrolytes are included in **Figure S13**).

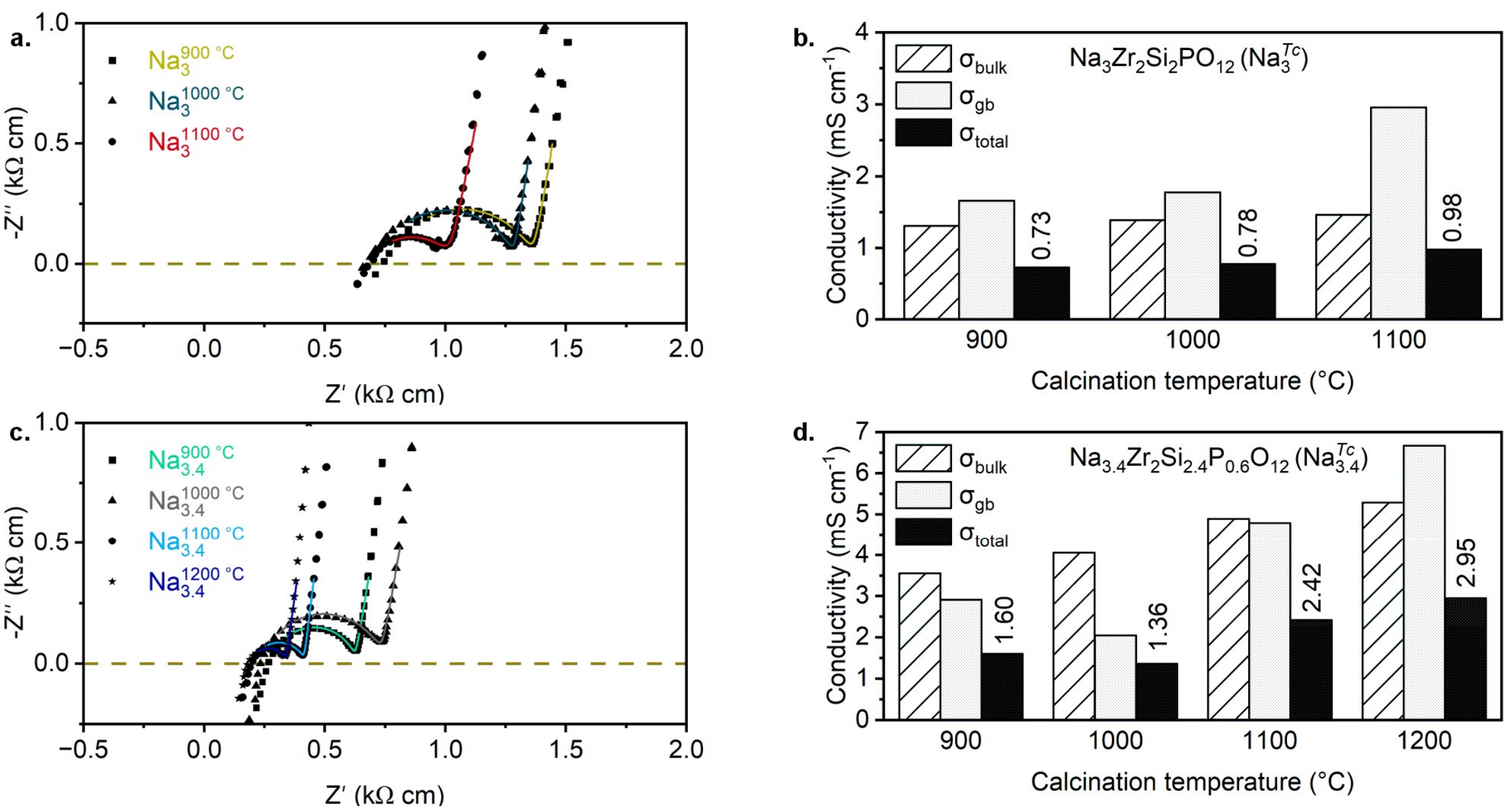

**Figure 6. Electrochemical characterization of NASICON electrolytes. a.** Potentiostatic electrochemical impedance spectroscopy (PEIS) measurements performed in a blocking-electrode configuration for the $Na_3$ electrolyte series using 200 nm sputtered Au electrodes. The colored lines correspond to the fitted curves. **b.** Ionic conductivity contributions (bulk, grain boundary) and total ionic conductivity computed from the fitting results for the $Na_3$ series. **c.** PEIS measurements and fitting for the $Na_{3.4}$ electrolyte series (the setup is analogous as in **a.**). **d.** Ionic conductivity contributions and total ionic conductivity computed from the fitting results for the $Na_{3.4}$ series. Numerical values for each contribution for both electrolyte compositions are reported in **Figure S12**.

High ionic conductivity alone is insufficient for practical operation, as compatibility with sodium metal electrodes determines the current densities that can be sustained without short circuiting. To evaluate Na compatibility under controlled mechanical conditions, symmetric Na | NASICON | Na cells using the $Na_3$ electrolytes are assembled using an inset cell under a set applied pressure. Among the $Na_3$ electrolytes, the material calcined at 1100 °C is selected for sodium symmetric-cell testing because of its superior performance. The $Na_3^{1100\,°C}$ exhibits a critical current density of 1.75 mA·cm$^{-2}$ at RT and ~10 MPa stack pressure (**Figure 7a**), and stable stripping/plating behavior under the investigated conditions (**Figure 7b**). The stable cycling behavior indicates that the dense microstructure produced through optimized calcination supports homogeneous sodium deposition and stripping and suppresses premature cell failure.

Overall, electrochemical measurements confirm that the influence of calcination and the amount of phase formation prior to sintering determine not only the degree of densification, but also grain-boundary transport. Therefore, the reaction state of the powder prior to sintering constitutes a key processing parameter governing the electrochemical performance of NASICON electrolytes.

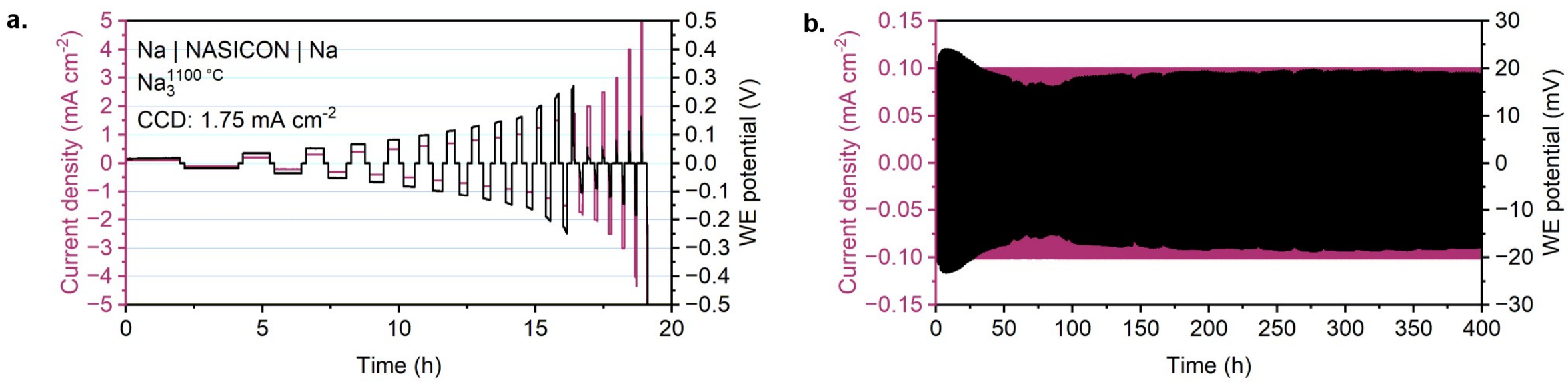

**Figure 7. Electrochemical stability and sodium plating/stripping performance of $Na_3$ NASICON electrolytes. a.** Critical current density (CCD) test of a symmetric Na | NASICON | Na cells employing a $Na_3^{1100\,°C}$ electrolyte at RT. The exchanged charge at each stripping or plating step is 0.1 mAh·cm$^{-2}$, with 10 min breaks at 0 V in-between steps, following the methodology proposed by Fuchs *et al.*[62] **b.** Galvanostatic cycling of a symmetric Na cell employing a $Na_3^{1100\,°C}$ electrolyte at room temperature. Stripping and plating steps are performed with a current density of 0.1 mA·cm$^{-2}$ and last 1 h each.

### 3.5 Sodium compatibility and effect of phosphorus compensation in $Na_{3.4}$ electrolytes

Despite exhibiting the highest ionic conductivity among the produced electrolytes and excellent densification, $Na_{3.4}^{1200\,°C}$ proved to be unstable in contact with metallic sodium. Although the electrolyte initially shows high conductivity and stable stripping/plating behavior, prolonged cycling results in a continuous decrease in cell impedance accompanied by increasingly unstable electrochemical response.

$Na_{3.4}^{1200\,°C}$ behaves coherently in contact with sodium only at very low current densities and for a limited time. **Figure 8a** highlights the shift between stable and unstable behavior of this electrolyte in a symmetric Na cell during CCD tests. Instead of exhibiting an increasing overpotential with increasing current density, the cell impedance continuously decreases. The same phenomenon is observed during long-term cycling at mild conditions (0.1 mA·cm$^{-2}$) (**Figure 8c**). After approximately 170 hours, the impedance contribution due to the electrolyte decreases (low-τ contribution in **Figure 8e**), leading to lower and unstable overvoltage during constant current cycling. Simultaneous decrease in bulk and grain-boundary resistance, as well as oscillating and incoherent voltage responses are inconsistent with interfacial contact loss or sodium dendrite formation (hard short circuit). Instead, the observations are consistent with the formation of soft short circuits in the electrolyte.[63,64] This phenomenon is linked to the formation of Na filaments or depositions within pores and grain boundaries, with the progressive development of a mixed ionic-electronic conduction mechanism within the electrolyte.[64,65]

The ICP-OES measurements presented in Section 3.3 reveal a phosphorus deficiency of approximately 20%, indicating significant phosphorus loss during synthesis. It is therefore hypothesized that this deviation in composition contributes to the observed premature failure of the SE. Although a comparable phosphorus deficiency is also detected for the $Na_3$ composition, instability is only observed for $Na_{3.4}$. This suggests that the consequences of P loss are stoichiometry-dependent. Because $Na_{3.4}$ contains a lower initial phosphorus content than $Na_3$, a similar relative phosphorus loss may promote a higher formation of phosphorus-deficient grain-boundary regions and secondary phases in $Na_{3.4}$ electrolytes.

To test this hypothesis, $Na_{3.4}$ powder containing a 20% P excess is synthesized with calcination performed at 1200 °C. This material is called $Na_{3.4+0.2P}^{1200\,°C}$. The material and electrochemical characterization of $Na_{3.4+0.2P}^{1200\,°C}$ follows the same logic as the other batches and the respective results are reported in the **Supporting Information**. In this section, the electrochemical performances of $Na_{3.4+0.2P}^{1200\,°C}$ and $Na_{3.4}^{1200\,°C}$ are compared. The P-compensated electrolyte shows markedly improved stability against sodium metal. The CCD response is characteristic of a stable solid electrolyte, with the overpotential increasing progressively with applied current density until failure starts to occur at 5.0 mA·cm$^{-2}$ (**Figure 8b**). This, in combination with the high measured RT conductivity of 3.82 mS·cm$^{-1}$, shows the superior performance of the $Na_{3.4+0.2P}^{1200\,°C}$ electrolytes not only compared to the non-compensated $Na_{3.4}$, but also to the $Na_3$ electrolytes (EIS fitting of the $Na_{3.4+0.2P}^{1200\,°C}$ SE is reported in **Figure S12**). The electrolyte sustains stable stripping/plating for over 400 h without the progressive decrease in impedance observed for the uncompensated material (**Figure 8d**). In this case, the changes in overpotential are associated with the evolution of the interface with Na (**Figure 8f**). DRT analysis shows how, in the case of the $Na_{3.4+0.2P}^{1200\ °C}$ electrolyte, the high-frequency impedance contribution due to the electrolyte response remains stable, while the low-frequency impedance contribution evolves (at higher τ values).

These results show how processing-induced compositional changes, even when they have only a limited influence on bulk ionic transport, can alter the long-term stability of NASICON electrolytes against sodium metal. Controlling P loss during synthesis is therefore essential not only for preserving the target composition but also for ensuring electrochemical robustness.

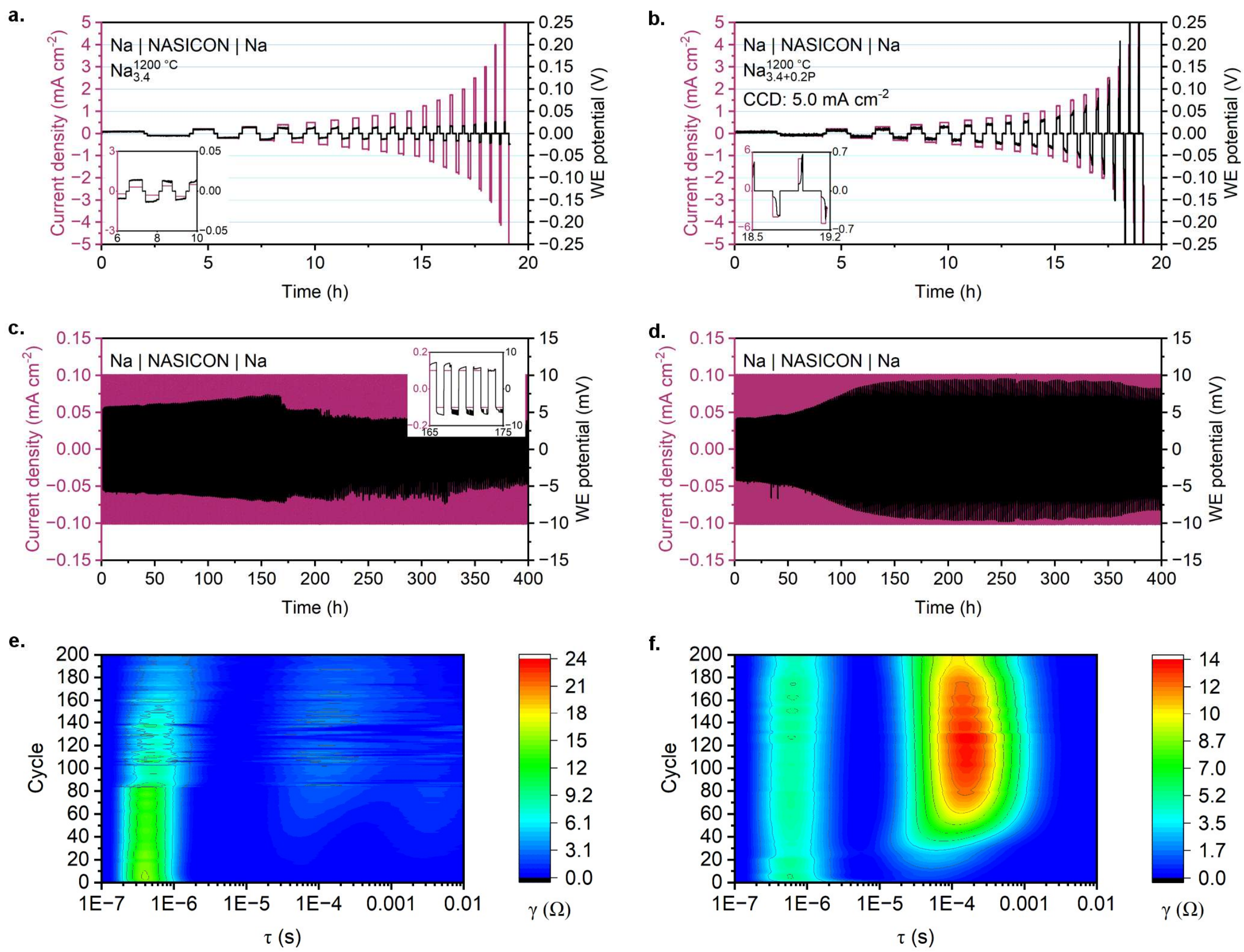

**Figure 8. Instability of Na3.4 NASICON and stabilization through phosphorus compensation. a-b.** CCD measurements at room temperature of symmetric Na | NASICON | Na cells assembled with $Na_{3.4}^{1200\ °C}$ and $Na_{3.4+0.2P}^{1200\ °C}$ electrolytes, showing that compensation for phosphorus loss during synthesis stabilizes the NASICON structure and improves the CCD and cycling stability of electrolytes. The inset in (**a.**) shows the onset of the abnormal impedance decrease at a constant current density, which marks the beginning of the unstable operation region. The inset in (**b.**) shows the onset of interfacial failure after the swift increase in interfacial impedance at higher

current densities. **c-d.** Galvanostatic cycling of symmetric Na cells employing $Na_{3.4}^{1200\,°C}$ and $Na_{3.4+0.2P}^{1200\,°C}$ electrolytes at RT. The inset in (**c.**) shows the beginning of the unstable cycling region for $Na_{3.4}^{1200\,°C}$, in which impedance of both bulk and grain boundary of the electrolyte start decreasing incoherently. **e-f.** Distribution of relaxation times (DRT) performed on PEIS measurements of $Na_{3.4}^{1200\,°C}$ (**e.**) and $Na_{3.4+0.2P}^{1200\,°C}$ (**f.**) electrolytes. $\tau$ is the relaxation time, and $\gamma(\tau)$ is the distribution of relaxation processes.

# 4. Conclusions

The effect of changing calcination conditions on NASICON densification and electrochemical properties is explored for the widely-used SSR synthesis of the SE. With calcination temperatures ranging from 900 °C to 1200 °C, powders with different amounts of NASICON phase prior to sintering produce electrolytes with different performances. The onsets of the high-temperature NASICON formation for $Na_3Zr_2Si_2PO_{12}$ and $Na_{3.4}Zr_2Si_{2.4}P_{0.6}O_{12}$ are identified. Compatibly with what previous studies showed, the latter stoichiometry reacts and densifies at higher temperatures.

These results show that a complete or near-complete NASICON phase formation before sintering leads to denser electrolytes, with lower grain-boundary impedance compared to reactive sintering. The most conductive electrolytes of each composition are indeed the materials that presented the highest NASICON phase content prior to sintering: $Na_{3}^{1100\,°C}$ and $Na_{3.4}^{1200\,°C}$, reaching grain-boundary conductivities of 2.95 and 6.67 mS·cm$^{-1}$, respectively. This positive effect is less pronounced for the bulk contribution of impedance but nonetheless leads to higher ionic conductivity.

Furthermore, the study shows how P loss during processing can be linked to a decrease in electrolyte stability and premature formation of soft short circuits in the electrolyte. The addition of an excess P amount to the raw powders leads to the stabilization of the SE in contact with Na electrodes. These stabilized electrolytes reach a critical current density of 5.0 mA·cm$^{-2}$ and an improved total conductivity of 3.82 mS·cm$^{-1}$ at RT.

These findings highlight the importance of processing conditions prior to sintering on the performance of the electrolytes. Decoupling phase formation from densification in NASICON production is an aspect that has not been systematically explored in literature but leads to large variations in the final properties of the material. While this study refers to a specific selection of raw materials, correct estimation of the onset of NASICON phase formation is expected to be of great importance to produce denser and better performing electrolytes also with other synthesis routes. Good control and a rational choice of synthesis parameters are therefore likely to play an important role not only in SSR synthesis, but also for other synthesis routes, like sol-gel or the solution-assisted SSR. These aspects need to be explored and understood deeper, to improve electrolyte performance but also reproducibility of results throughout literature.

## Acknowledgments

The authors thank: Philipp Adelhelm (Humboldt-Universität zu Berlin, Helmholtz-Zentrum Berlin), Dominik Al-Sabbagh (BAM), Alain Cerny (BAM), Franziska Lindemann (BAM), Anastasia May (BAM), Stefan Reinsch (BAM), Luise Sander (BAM), Alexander Seidel (BAM, Humboldt-Universität zu Berlin) and Alice Guo Tao (BAM, UC Berkeley) for fruitful discussions and/or assistance in the labs. This work was supported by the Bundesministerium für Forschung, Technologie und Raumfahrt (BMFTR/BMBF) under grants 03XP0516, 03XP0462 (FORBATT project "TomoFestBattLab"), and 03XP0525C (project "NATTER"). Additional funding was provided by the School of Analytical Sciences Adlershof (SALSA, STF26-02) and the Fonds der Chemischen Industrie im Verband der Chemischen Industrie e. V. (grant 661737). The project in which this work was carried out is part of the Berlin Battery Lab (BBL).

## Data availability

The data generated and analyzed in this study has been deposited in Zenodo and is openly available at https://doi.org/10.5281/zenodo.22708928.

## Conflict of Interest

The authors declare that they have no conflict of interest.

## Bibliography

[1] G. Graeber, V. S. Thatipamula, *MIT SPR* **2022**, *3*.
[2] Q. Ma, F. Tietz, *ChemElectroChem* **2020**, *7*, 2693.
[3] A. J. Bard, L. R. Faulkner, *Electrochemical methods*, 2nd ed., Wiley, New York, **2001**.
[4] Y. Wang, J. Liu, J. Wang, C. Sun, *Inorg. Chem. Front.* **2025**, *12*, 6011.
[5] H. Liu, X.-B. Cheng, J.-Q. Huang, S. Kaskel, S. Chou, H. S. Park, Q. Zhang, *ACS Materials Lett.* **2019**, *1*, 217.

[6] J. Baller, A. Hilger, N. Qi, C. Morini, A. Cornelio, A. Remhof, M. Osenberg, I. Manke, J. Moosmann, F. Beckmann, G. Graeber, *Adv Funct Materials* **2026**, *36*, e23169.
[7] M. P. Fertig, K. Skadell, M. Schulz, C. Dirksen, P. Adelhelm, M. Stelter, *Batteries & Supercaps* **2022**, *5*, e202100131.
[8] Z. Deng, G. Sai Gautam, S. K. Kolli, J.-N. Chotard, A. K. Cheetham, C. Masquelier, P. Canepa, *Chem. Mater.* **2020**, *32*, 7908.
[9] H. Y.-P. Hong, *Materials Research Bulletin* **1976**, *11*, 173.
[10] J. B. Goodenough, H. Y.-P. Hong, J. A. Kafalas, *Materials Research Bulletin* **1976**, *11*, 203.
[11] L. Liu, J. Su, X. Zhou, D. Liang, Y. Liu, R. Tang, Y. Xu, Y. Jiang, Z. Wei, *Materials Today Chemistry* **2023**, *30*, 101495.
[12] Q. Ma, T. Ortmann, A. Yang, D. Sebold, S. Burkhardt, M. Rohnke, F. Tietz, D. Fattakhova-Rohlfing, J. Janek, O. Guillon, *Advanced Energy Materials* **2022**, *12*, 2201680.
[13] L. Liu, D. Liang, X. Zhou, Y. Liu, J. Su, Y. Xu, J. Peng, *J Mater Sci* **2022**, *57*, 11774.
[14] M. Iordache, A. Oubraham, I. Petreanu, C. Sisu, S. Borta, C. Capris, A. Soare, A. Marinoiu, *Materials* **2024**, *17*, 823.
[15] R. Thirupathi, R. P. Srivastava, B. Patankar, S. Bhattacharyya, M. Aman, S. Sharma, S. Omar, *Chem. Commun.* **2025**, *61*, 10931.
[16] R. Fuentes, *Solid State Ionics* **2001**, *140*, 173.
[17] A. Jalalian-Khakshour, C. O. Phillips, L. Jackson, T. O. Dunlop, S. Margadonna, D. Deganello, *J Mater Sci* **2020**, *55*, 2291.
[18] S. Naqash, D. Sebold, F. Tietz, O. Guillon, *J Am Ceram Soc* **2019**, *102*, 1057.
[19] B. Xun, J. Wang, H. R. Arindra, K. Hayashi, *Adv Eng Mater* **2026**, *28*, e202501808.
[20] J. Wolfenstine, W. Go, Y. Kim, J. Sakamoto, *Ionics* **2023**, *29*, 1.
[21] S.-P. Guo, J.-C. Li, Q.-T. Xu, Z. Ma, H.-G. Xue, *Journal of Power Sources* **2017**, *361*, 285.
[22] A. Marinoiu, M. Iordache, A. Tiliakos, *Revue Roumaine de Chimie* **2022**, *67*, 63.
[23] E. Dashjav, M. Bhardwaj, M.-T. Gerhards, Q. Ma, K. Wätzig, C. Baumgärtner, D. Wagner, A. Lowack, M. Kusnezoff, F. Tietz, *ACS Appl. Energy Mater.* **2025**, *8*, 11373.
[24] J. G. Pereira Da Silva, M. Bram, A. M. Laptev, J. Gonzalez-Julian, Q. Ma, F. Tietz, O. Guillon, *Journal of the European Ceramic Society* **2019**, *39*, 2697.
[25] B. Hitesh, A. Sil, *J Am Ceram Soc.* **2023**, *106*, 6743.
[26] A. Tiwari, D. Meghnani, R. Mishra, R. K. Tiwari, A. Patel, R. K. Singh, *Journal of Power Sources* **2023**, *580*, 233365.
[27] C.-H. Yang, T.-H. Lung, W.-R. Liu, *Journal of Alloys and Compounds* **2026**, *1050*, 185491.
[28] Y. Li, Z. Sun, C. Sun, H. Jin, Y. Zhao, *Ceramics International* **2023**, *49*, 3094.
[29] A. S. Peretti, E. D. Spoerke, M. E. Ureña, I. D. Dyer, P. A. Salinas, Mark. A. Rodriguez, P. S. Mantos, J. N. Williard, L. J. Small, *J Am Ceram Soc.* **2026**, *109*, e70195.
[30] Y. Xing, Y. Li, C. Zhang, *Solid State Ionics* **2021**, *373*, 115811.
[31] B. Wei, S. Huang, X. Wang, M. Liu, C. Huang, R. Liu, H. Jin, *Energy Environ. Sci.* **2025**, *18*, 831.
[32] C. Vakifahmetoglu, L. Karacasulu, *Current Opinion in Solid State and Materials Science* **2020**, *24*, 100807.
[33] L. Ran, A. Baktash, M. Li, Y. Yin, B. Demir, T. Lin, M. Li, M. Rana, I. Gentle, L. Wang, D. J. Searles, R. Knibbe, *Energy Storage Materials* **2021**, *40*, 282.
[34] T. Takahashi, K. Kuwabara, M. Shibata, *Solid State Ionics* **1980**, *1*, 163.
[35] S. Guan, J. Lu, Y. Li, D. Xie, C. Zhuang, W. Zhang, *Ceramics International* **2025**, *51*, 1172.
[36] J. Luo, G. Zhao, W. Qiang, B. Huang, *J Am Ceram Soc* **2022**, *105*, 3428.

[37] Y. Ruan, S. Song, J. Liu, P. Liu, B. Cheng, X. Song, V. Battaglia, *Ceramics International* **2017**, *43*, 7810.
[38] A. Chakraborty, R. Thirupathi, S. Bhattacharyya, K. Singh, S. Omar, *Journal of Power Sources* **2023**, *572*, 233092.
[39] Q. Zhang, Q. Zhou, Y. Lu, Y. Shao, Y. Qi, X. Qi, G. Zhong, Y. Yang, L. Chen, Y.-S. Hu, *Engineering* **2022**, *8*, 170.
[40] Z. Khakpour, *Electrochimica Acta* **2016**, *196*, 337.
[41] S. K. Pal, R. Saha, G. V. Kumar, S. Omar, *J. Phys. Chem. C* **2020**, *124*, 9161.
[42] Y. Shao, G. Zhong, Y. Lu, L. Liu, C. Zhao, Q. Zhang, Y.-S. Hu, Y. Yang, L. Chen, *Energy Storage Materials* **2019**, *23*, 514.
[43] J. M. Naranjo-Balseca, C. S. Martínez-Cisneros, B. Pandit, A. Várez, *Journal of the European Ceramic Society* **2023**, *43*, 4826.
[44] V. M. Pratiwi, A. A. Wibowo, Widyastuti, H. Purwaningsih, F. A. Maulana, *MSF* **2019**, *964*, 168.
[45] V. M. Pratiwi, L. Noerochiem, Widyastuti, H. Purwaningsih, D. Susanti, F. A. Maulana, Surabaya, Indonesia, **2021**, *2384*, p. 050007.
[46] J. Lee, C. Chang, Y. I. Lee, J. Lee, S. Hong, *Journal of the American Ceramic Society* **2004**, *87*, 305.
[47] H. Leng, J. Huang, J. Nie, J. Luo, *Journal of Power Sources* **2018**, *391*, 170.
[48] M. Samiee, B. Radhakrishnan, Z. Rice, Z. Deng, Y. S. Meng, S. P. Ong, J. Luo, *Journal of Power Sources* **2017**, *347*, 229.
[49] D. N. Karmalkar, D. Dutta, A. K. Bera, S. M. Yusuf, B. Pahari, *J. Phys. Chem. C* **2024**, *128*, 768.
[50] P. Kumar, S. S. Jadaun, S. Singh, G. Ratan, A. K. Panwar, *Materials Research Bulletin* **2027**, *205*, 114348.
[51] Y. Li, Z. Sun, H. Jin, Y. Zhao, *Batteries* **2023**, *9*, 252.
[52] L. Shen, J. Yang, G. Liu, M. Avdeev, X. Yao, *Materials Today Energy* **2021**, *20*, 100691.
[53] A. Tiliakos, M. Iordache, A. Marinoiu, *Applied Sciences* **2021**, *11*, 8432.
[54] X. Feng, Z. Luo, T. Wu, J. Yang, H. Liang, Y. Li, *Inorganic Chemistry Communications* **2025**, *182*, 115609.
[55] S. Hu, J. Chang, Y. Kong, X. Liu, R. Yang, J. Wang, H. Sun, K. Zhang, G. Hu, W. Hu, J. Zhang, K. Hong, *PAC* **2025**, *19*, 36.
[56] E. Quérel, I. D. Seymour, A. Cavallaro, Q. Ma, F. Tietz, A. Aguadero, *J. Phys. Energy* **2021**, *3*, 044007.
[57] M. Bay, M. Wang, R. Grissa, M. V. F. Heinz, J. Sakamoto, C. Battaglia, *Advanced Energy Materials* **2020**, *10*, 1902899.
[58] A. A. Coelho, *J Appl Crystallogr* **2018**, *51*, 210.
[59] V. Yrjänä, *JOSS* **2022**, *7*, 4808.
[60] A. Loutati, Y. J. Sohn, F. Tietz, *ChemPhysChem* **2021**, *22*, 995.
[61] T. Ortmann, S. Burkhardt, J. K. Eckhardt, T. Fuchs, Z. Ding, J. Sann, M. Rohnke, Q. Ma, F. Tietz, D. Fattakhova-Rohlfing, C. Kübel, O. Guillon, C. Heiliger, J. Janek, *Advanced Energy Materials* **2023**, *13*, 2202712.
[62] T. Fuchs, C. G. Haslam, F. H. Richter, J. Sakamoto, J. Janek, *Advanced Energy Materials* **2023**, *13*, 2302383.
[63] M. J. Counihan, K. S. Chavan, P. Barai, D. J. Powers, Y. Zhang, V. Srinivasan, S. Tepavcevic, *Joule* **2024**, *8*, 64.
[64] C. Wang, Y. He, P. Zou, Q. He, J. Li, H. L. Xin, *J. Am. Chem. Soc.* **2025**, *147*, 19084.
[65] Q. Li, A. Chen, D. Wang, Z. Pei, C. Zhi, *Joule* **2022**, *6*, 273.